\documentclass[prl,aps,preprint]{revtex4}

\usepackage{graphicx}

\begin{document}

\title{Manifestation of interdot spin tunneling effects in ordered arrays of quantum dots:   An anomalous magnetization of $In_{1-x}Mn_xAs$ }

\author{Sergei  Sergeenkov$^{1}$ and Marcel Ausloos$^{2,3}$}

\affiliation{$^{1}$Departamento de F\'{\i}sica, Universidade Federal da Para\'{\i}ba, 58051-970 Jo\~{a}o Pessoa, PB, Brazil\\
$^{2}$Group for Research on Applications of Physics in Economy and Sociology (GRAPES), rue de la Belle Jardiniere, 483/021, B-4031 Liege Angleur, Belgium\\ 
$^{3}$School of Management, University of Leicester, University Road, Leicester LE1 7RH, United Kingdom\\}

\date{\today}

\begin{abstract}

The anomalous low-temperature behavior of the spontaneous magnetization  of ordered arrays of $In_{1-x}Mn_xAs$ quantum dots is discussed. It is shown that the experimental results can be well understood, within a mean field approximation,  assuming  collective response through rather strong spin tunneling processes between neighbouring dots.\\
 
{\bf Keywords}: Quantum Dots; Ordered Array; Magnetization; Spin Tunnelling 
 
\end{abstract}

\maketitle

\section{1. Introduction}

Man-made nano-structures called "quantum dots" (QDs) are about 25 years old \cite{a1}. The manufacturing  consists in quantum confining a few electrons  in all three spatial
dimensions. A common way of fabrication of these artificial atoms is to restrict the two-dimensional electron gas in a semiconductor heterostructure laterally by electrostatic gates, or vertically by etching techniques. The  study  of such nanostructure systems has been  quite developed in recent years. Much theoretical insight has been gained  concerning the electronic ground-state structure and  through many-body physics  considerations  \cite{a2}.   A review of the statistical theory of QDs  with emphasis on chaotic or diffusive electron dynamics, was  provided in \cite{a3}. 

Moreover, recent investigations have reported antiferromagnetic (AF) correlations between two single-level coupled QDs, in competition with Kondo correlations, but 
a regime with ferromagnetism has also been found, if more than one level per dot is active \cite{a4}. Furthermore, in recent years,  systems  have been realized in which individual electrons can be trapped and their quantum properties can be studied, thus avoiding unnecessary ensemble averaging \cite{a5}.  However, single "site" properties, even within a mean field approximation,  may depend on  (or be influenced by) electronic resonant states \cite{a6}, in other words in the present QD cases, spin tunneling. These considerations are  of interest for a simple approach of an apparently anomalous phenomenon discussed here below.
 
The artificially prepared ordered arrays of  QDs based on the ternary $In_{1-x}Mn_xAs$ alloys continue to attract a significant attention due to their numerous potential applications (see, e.g., \cite{b1,b2,b3,b4} and further references therein).  

In this Letter,  we briefly discuss a possible manifestation of spin tunneling effects between closely packed dots in the temperature dependence of the spontaneous magnetization $M(T)$ observed in some of these arrays.   The data is presented in Section 2, leading to  the consideration that the anomalous behavior is due to intra- and inter-QD spin tunneling. In Section 3,  serving as a conclusion, we question whether other data should be re-examined at low temperature, and suggest theoretical improvements based on the present findings.

\section{2. Data and Analysis}

The description of the production and the arrays characterization, e.g. structural properties, of the ternary $In_{1-x}Mn_xAs$ alloys will be fully discussed elsewhere \cite{b5}. Nevertheless, let it be known that the  self-assembled ferromagnetic (FM) $In_{1-x}Mn_xAs$  QDs (with $Mn$ concentrations $x$ in the range of $0.01<x<0.3$) were grown by molecular beam epitaxy following an original method to order QDs by using a non-magnetic $GaAs(100)$ template~\cite{b4}. To measure the temperature variation of the low-field magnetization with high precision, a homemade magnetometer was used \cite{b6}. A typical  $M(T)$ curve for $x=0.25$ is shown in Fig.1; the measurements details will be presented elsewhere  \cite{b5}. 

A major point needs to be appreciated: it is worth noting that in addition to a strong FM behavior (with the well-defined Curie temperature around $300K$) there is a clear evidence in favor of a second transition (around an inflection point $T_0=69K$). Given a rather small distance between dots within an array  \cite{b5}($d=2nm$), it is quite reasonable to assume that  a special collective effect induces this phase transition. We propose that it is due to (Mn) spins tunneling between neighboring dots. Some analogy can be taken from the intragrain-intergrain currents on flux profile in granular superconducting ceramics \cite{c1,c2,c3,c4}.  
By attributing the high-temperature region (above $T_0$, see Fig.2) to the intradot magnetization $M_d$ (with the Curie temperature $T_{Cd}$) and the low-temperature region (below $T_0$) to the interdot  magnetization $M_t$ (with the Curie temperature $T_{Ct}$),  one can successfully fit the experimental data  using the following expressions 
\begin{equation}
M_d(T)=M_{sd}\tanh\left[\sqrt{\left(\frac{T_{Cd}}{T}\right)^2-1}\right]+
M_{md}(0)\left[1-A_d\left(\frac{T}{T_{Cd}}\right)^{3/2}\right]
\end{equation}
and
\begin{equation}
M_t(T)=M_{st}\tanh\left[\sqrt{\left(\frac{T_{Ct}}{T}\right)^2-1}\right]+
M_{mt}(0)\left[1-A_t\left(\frac{T}{T_{Ct}}\right)^{3/2}\right]
\end{equation}
above and below the inflection point $T_0$, respectively. Here, $M_{sd}$ and $M_{st}$ are the corresponding saturation magnetizations. The first terms in the rhs of  Eqs.(1) and (2) present  analytical (approximate) solution of the well-known Curie-Weiss mean-field equation for spontaneous magnetization valid for all temperatures (\cite{c3,c4}), while the second terms account for the Bloch (magnon) contributions. Fig.2 presents the best fits of the experimental data according to Eqs.(1) and (2)  for the fit parameters: $M_{sd}=0.87M_t(0)$, $M_{st}=0.97M_t(0)$, $M_{md}(0)=0.03M_t(0)$,  $M_{mt}(0)=0.03M_t(0)$, $A_d=A_t=0.26$, $T_{Cd}=299K$,  $T_{Ct}=143K$, and $M_t(0)=2.25\times 10^{-5}emu$.

\begin{figure}
\centerline{\includegraphics[width=11cm]{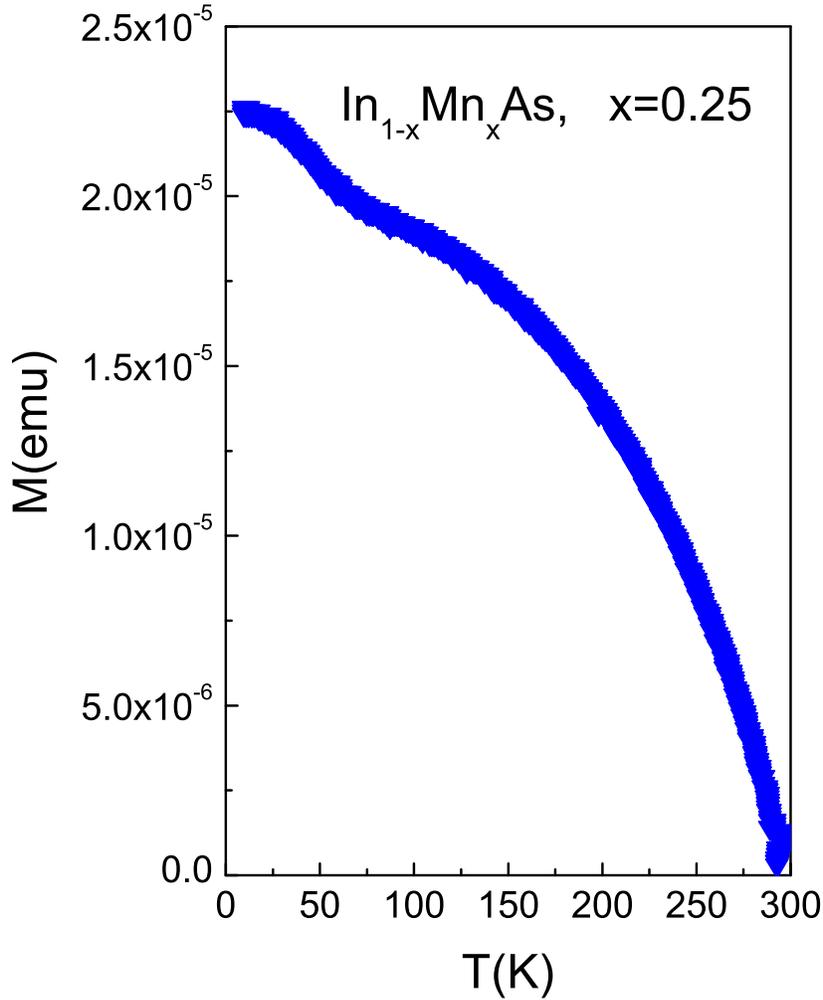}}
\caption{Temperature dependence of the measured magnetization for an array of $In_{1-x}Mn_xAs$ dots with $x=0.25$ (courtesy of V.A.G. Rivera \cite{b5}).} \label{fig:fig1}
\end{figure}

\begin{figure}
\centerline{\includegraphics[width=11cm]{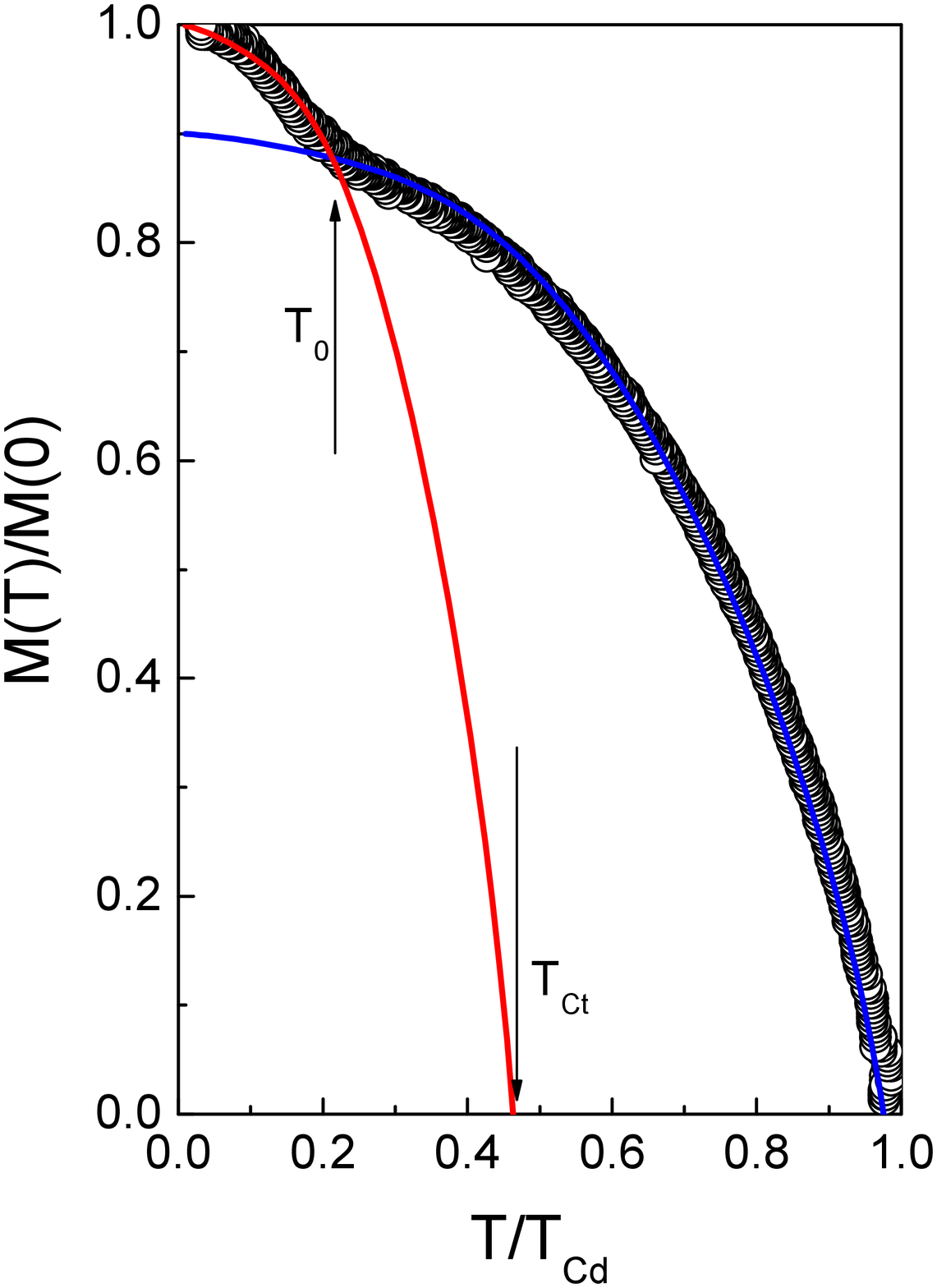}}
\caption{Fit of the normalized experimental data (shown in Fig.1). The solid blue and red lines are the intradot and interdot contributions according to Eq.(1) and Eq.(2), respectively.}
\label{fig:fig2}
\end{figure}

Recall \cite{d1} that FM in single $In_{1-x}Mn_xAs$ dots is most likely due to carriers mediated RKKY type exchange mechanism with a local energy $-J_{ij}S_iS_j$ between Mn spins at sites $i$ and $j$, where the spin exchange energy $J_{ij}$ is related to the intradot Curie temperature as follows $T_{Cd}(x)=xS(S+1)zJ_{ij}/3$. Here, $S$ and $x$ are the spin and concentration of Mn atoms, and $z$ is the number of nearest neighbors. Using $S=5/2$, $z=2$ and the experimentally found $T_{Cd} =299K$, we obtain $J_{ij}=16meV$ for a reasonable estimate of the spin exchange coupling energy for a single dot.  In turn, this value allows us to estimate the model parameter $A_d$ entering the Bloch law given by the second term in Eq.(1), namely \cite{d2} $A_d =(k_BT_{Cd}/J_{ij}S)^{3/2}=0.25$ in good agreement with the fitting value used for this parameter. Turning to the discussion of the interdot contribution $M_t(T)$, it is reasonable to assume that the FM behavior between the nearest dots is still governed by the same RKKY type exchange mechanism strongly modulated by the interdot tunneling probability, that is $J_t=\exp(-d/\xi)J_{ij}$ where $d$ is the distance between neighboring dots, $\xi=\hbar /\sqrt{2mU}$  is a characteristic length with $U$ being the barrier height, and $m$ the carrier mass. 
As a result, the interdot Curie temperature $T_{Ct}$ is related to the intradot Curie temperature $T_{Cd}$ as follows $T_{Ct} =\exp(-d/\xi)T_{Cd}$. This means that $A_t=(k_BT_{Ct}/J_tS)^{3/2}=A_d$, as expected. Furthermore, using $d=2nm$ along with the found values of the Curie temperatures ($T_{Cd}=299K$ and $T_{Ct}=143K$), we obtain $\xi=2.5nm$ for an estimate of the characteristic length which corresponds to the barrier height of $U=6meV$ (assuming free electron mass for $m$). Notice also that this energy, in turn, corresponds to a characteristic temperature $T_0=U/k_B= \hbar ^2/2m\xi ^2=69K$ which remarkably correlates with the inflection point between intradot and interdot contributions observed experimentally (Cf. Figs. 1 and 2). 

\section{3. Conclusions}

Finally,  as a conclusion, let us note that this (rather good) description of highly doped films by such simple and easily understandable expressions,  Eqs.(1) and (2),  suggests  the existence of a coherent response from all the dots forming an array, (despite a somewhat disordered distribution of Mn spins  as seen via structural measurements ~\cite{b4,b5}).   We stress that the "low critical temperature" value $T_{Ct}$ is masked by the fluctuations and by the high temperature phase, but can be truly obtained from the shoulder temperature $T_0$, which is not  a true critical temperature. 
This anomaly might be relevant in analyzing other static \cite{d3} and also transport  properties at (magnetic) phase transitions  \cite{d4} in QDs systems. It can be also easily  imagined that renormalization group approaches \cite{d5} would lead to a better estimate of the exponents  in Eqs. (1)--(2),  thereby in refining the given values of the parameters.

\vskip0.5cm
The authors are indebted to V.A.G. Rivera for making available the experimental data prior to publication. This work has been partially financially supported by the Brazilian agency FAPESQ (DCR-PB).

\end{document}